# Evidence of Micron-Scale Ion Damage in (010), (110), and (011) β-Ga$_2$O$_3$ Epitaxial Layers

Carl Peterson,[1,a),*] Chinmoy Nath Saha,[1,*] Yizheng Liu,[1] James S. Speck,[1] and Sriram Krishnamoorthy[1,a)]

[1]*Materials Department, University of California Santa Barbara, Santa Barbara, California, 93106, USA*

\* Authors contributed equally to this work

___________________________

a) Author to whom correspondence should be addressed.  Electronic mail: carlpeterson@ucsb.edu and sriramkrishnamoorthy@ucsb.edu

We report on the experimental observation of up to 11.5 µm deep charge depletion in (010), (110), and (011) *β*-Ga$_2$O$_3$ epitaxial layers due to ion damage from sputtering and inductively coupled plasma (ICP) etching processes whereas charge depletion in (001) *β*-Ga$_2$O$_3$ epitaxial layers was minimal. The orientation-dependent reduction in CV-measured charge density was first observed in NiO$_x$ reactively sputtered heterojunction p-n diodes (HJDs). When compared to reference low-damage Schottky barrier diodes (SBDs), the sputtered HJDs showed a 9.4× increase in the specific on resistance (R$_{on,sp}$) and 85% reduction in net donor concentration ($N_D - N_A$) at zero bias for sputter-damaged HJDs on (010) epitaxial layers whereas HJDs on (001) remained unchanged. Similarly, sputtered SiO$_2$ caused a reduction of $N_D - N_A$ 11.5 µm deep into the (010) material. Next, SBDs were fabricated on *β*-Ga$_2$O$_3$ surfaces previously etched via a BCl$_3$ based ICP process and compared to SBDs on un-etched surfaces. The (010) SBDs on etched surfaces exhibited a 7.7× increase in R$_{on,sp}$ and a 91% reduction in $N_D - N_A$ at zero bias where the (001) etched diodes exhibited little change. Additionally, (110) and (011) diodes fabricated on ICP damaged surfaces also saw a ~82% reduction in $N_D - N_A$ at zero bias, indicating (110) and (011) are also susceptible to ion damage. Damage in the (010), (110), and (011) diodes is potentially caused by energetic ions that travel into the open channels present along the [010] direction and create compensating point defects which could potentially diffuse further.

Beta Gallium Oxide ($\beta$-Ga$_2$O$_3$) is an ultra-wide band gap (UWBG) semiconductor which has shown promise for creating energy-efficient power electronic devices.[1] A major salient feature of $\beta$-Ga$_2$O$_3$ in comparison with GaN, SiC, and other emerging UWBG semiconductors is the availability of melt-grown conductive and insulating bulk substrates with controlled impurity doping.[2–8] The melt-growth method enables low extended defect densities in the substrates and epitaxial layers, which can improve device reliability and yield, as well possibly offer low final device costs due to large area scalability. Additionally, the combination of $\beta$-Ga$_2$O$_3$'s bandgap (4.6 - 4.9 eV), predicted critical electric field strength of 8 MV/cm, mobility of 200 cm$^2$/Vs,[9] and availability of shallow dopants allows for $\beta$-Ga$_2$O$_3$ to be the best-in-class material among its WBG and UWBG peers when looking at figure of merit (FOM) analysis.[10,11] Recently, it has been reported experimentally that $\beta$-Ga$_2$O$_3$ can support even higher critical electric fields than predicted, up to 12.9 MV/cm,[12] making the UWBG semiconductor an even more appealing option for creating the next generation of electronic devices capable of operating in the multi-kV-class regime.[13,14]

When fabricating high voltage devices, popular techniques such as dry etching, ion implantation, and sputtering processes can expose $\beta$-Ga$_2$O$_3$ to energetic ions, potentially damaging the semiconductor and degrading device performance.[15] Despite low-damage alternatives such as heated H$_3$PO$_4$ wet etching,[16,17] HCl gas etching,[18] Ga etching,[19–23] and atomic layer deposition, potentially damaging techniques such as inductively coupled plasma (ICP) etching and sputtering are commonly performed to create fins and deposit dielectrics layers, respectively. Fins and dielectrics are both essential features for high-performance power devices such as vertical FinFETs,[14,24–27] lateral FinFETs,[28–32] and trench diodes.[33–39] However, the performance of devices with dry etched (010)-like fin sidewalls was lower than devices with dry etched (100)-like sidewalls,[40–43] suggesting that there is anisotropic crystal damage in $\beta$-Ga$_2$O$_3$ from ion processes. In this work, we experimentally demonstrate the extent of ion damage in the (001), (010), (110), and (011) $\beta$-Ga$_2$O$_3$ crystal orientations from both sputtering and ICP etching.

The first method for inducing ion damage into $\beta$-Ga$_2$O$_3$ epitaxial films was using RF sputtering. For this experiment, two 7 μm thick films with a controlled net donor concentration ($N_D - N_A$) = 1 × 10$^{16}$ cm$^{-3}$ films were used. The first film was an (001) halide vapor phase epitaxy (HVPE)-grown epilayer purchased commercially from Novel Crystal Technology Inc., Japan (NCT) and the second was an (010) epilayer grown using metal organic chemical vapor deposition (MOCVD) at UCSB. The MOCVD growth conditions for the 7 μm thick TMGa drift layer were optimized and reported in a previous work.[44] On the (010) and (001) films, Schottky barrier diodes (SBDs) and NiO$_x$ heterojunction p-n diodes (HJDs) were co-fabricated. The sputtered NiO$_x$ process is of particular interest as the NiO$_x$ HJDs have shown impressive performance with multi-kV-class breakdown voltages and reverse bias leakage less than the measurement equipment noise limit.[13,45–52] Device processing began with lithographic patterning of circular diode structures. A 20 nm p$^-$ layer of NiO$_x$ followed by a 20 nm p$^+$ layer of NiO$_x$ was then reactively sputtered using a Ni target and a 150 W O$_2$/Ar plasma without breaking vacuum. The resistivity of the p$^+$ layer was controlled by increasing the O$_2$/Ar ratio from the p$^-$ layer[51,53]. The samples were then immediately loaded into an electron-beam (e-beam) evaporator, and a self-aligned Ni/Au/Ni stack was deposited as an Ohmic contact to the NiO$_x$ layer. Lift-off processing was performed to finalize the HJD device structure. Circular Ni/Au/Ni SBDs were then fabricated on the remaining sample surface which was protected by photoresist during the HJD fabrication. Finally, a Ti/Au Ohmic contact was deposited on the back side of the samples using e-beam evaporation. The device schematics for both the HJD and SBD devices are shown in FIG. 1(a).

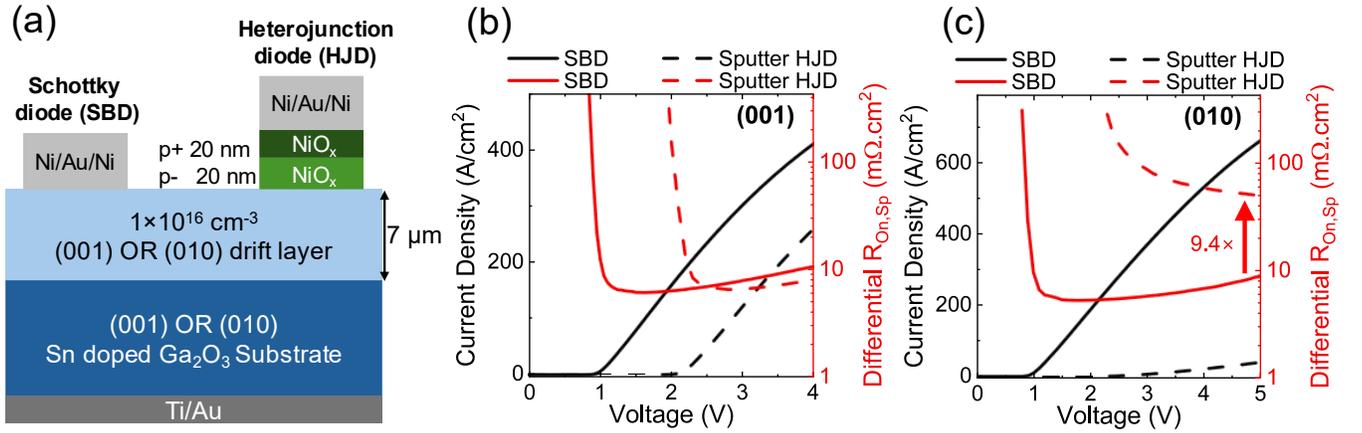

FIG. 1. (a) Device schematic of the reference SBDs and NiO$_x$ HJDs co-fabricated on (001) and (010) β-Ga$_2$O$_3$ drift layers. (b) J-V and differential R$_{on,sp}$ plots for (001) SBD and ion-damaged HJD diodes (c) J-V and differential R$_{on,sp}$ plots for (010) SBD and ion-damaged HJD diodes, showing a 9.4× increase in R$_{on,sp}$ for sputter damaged (010) devices.

Current density vs. voltage (J-V) measurements were performed on a Keithley 4200 parameter analyzer. FIG. 1(b) shows the J-V and differential specific on resistance (R$_{on,sp}$) characteristics of the SBD and HJD diodes fabricated on the (001) orientation. The J-V characteristics show a turn-on voltage (V$_{on}$) of 1 V and 2.2 V for the SBD and p-n HJD respectively, which is similar to NiO$_x$ HJD results from the literature.[50] The R$_{on,sp}$ plots show comparable minimum values of 6.14 and 6.57 mΩ.cm$^2$ for the SBD and HJD devices respectively. For the diodes on (010) material, however, the J-V and R$_{on,sp}$ results in FIG. 1(c) show a markedly lower current density and a 9.4× increase (from 5.30 to 49.91 mΩ.cm$^2$) in the minimum R$_{on,sp}$ for the HJD compared to the SBD. This signifies the 150 W NiO$_x$ sputtering caused significant ion damage only in the (010) material.

High-voltage capacitance-voltage (C-V) measurements were performed on a Keysight B1505A parameter analyzer to further quantify the ion damage in the (010) and (001) epilayers by extracting the net donor concentration ($N_D - N_A$) vs. depth profile. C-V on the (001) epilayer showed nearly identical $N_D - N_A$ profiles for both the HJD and SBD diodes (FIG. 2(a)), indicating minimal change in $N_D - N_A$ after sputtering. In contrast, for the (010) epilayer, FIG. 2(b) shows the sputtered HJD had a significant decrease in $N_D - N_A$ when compared to the SBD. The net donor concentration in the entire 7 μm (010) epilayer is below the reference $1\times10^{16}$ cm$^{-3}$, with an 85% reduction in $N_D - N_A$ at zero bias due to the 150 W NiO$_x$ sputtering process. Since the reduction in donor concentration observed in FIG. 2(b) would cause an increase in $R_{on,sp}$, results from J-V and C-V are consistent for the (010) and (001) diodes.

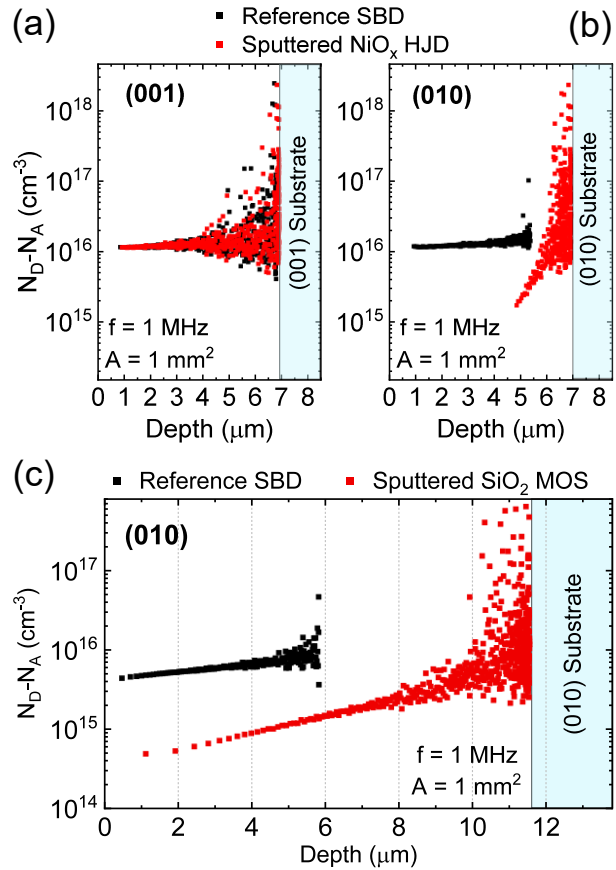

FIG. 2. $N_D - N_A$ vs. depth profiles from high-voltage C-V of reference SBD and sputtered HJD diodes on 7 μm thick $1\times10^{16}$ cm$^{-3}$ doped (a) (001) epitaxial layers and (b) (010) epitaxial layers. (c) $N_D - N_A$ vs. depth profiles of reference SBD and sputtered SiO$_2$ MOSCAPs, showing sputtered ion damage up to 11.5 microns into the (010) material.

To further confirm that sputtering was the cause of charge depletion and not NiO$_x$ itself, SiO$_2$ metal oxide semiconductor capacitors (MOSCAPs) were fabricated on an additional (010) sample. For this experiment, an 11.5 μm (010) epilayer was grown on a conductive Sn-doped (010) β-Ga$_2$O$_3$ substrate via MOCVD using the same growth conditions as above,[44] with a lower intended doping of $5\times10^{15}$ cm$^{-3}$. First, reference Ni/Au SBDs were created via e-beam evaporation and optical lithography. Next, a 380 nm SiO$_2$ layer was deposited by reactive sputtering at 245 W using an elemental Si target. The SiO$_2$ was then lifted off on top of the existing SBDs. Finally, Ni/Au circular metal contacts were deposited on top of the SiO$_2$ to create the MOSCAPs and a backside Ti/Au Ohmic contact was deposited to complete the vertical devices. In FIG. 2(c) the data for the reference SBDs shows a ~$5\times10^{15}$ cm$^{-3}$ $N_D - N_A$ profile whereas the sputter damaged MOSCAP structures showed a significant decrease in $N_D - N_A$ all the way through the 11.5 μm film, with an 89% reduction of $N_D - N_A$ at zero bias. Depletion of charge in both the sputtered NiO$_x$ HJDs and sputtered SiO$_2$ MOSCAPs confirms that the (010) orientation is susceptible to damage from various sputtering processes, not just NiO$_x$.

In addition to sputter, an experiment using ICP to damage the material was performed to further isolate energetic ions as the root cause for charge depletion and material damage. In this experiment, reference SBDs were co-fabricated along with SBDs on ICP damaged material. Devices were fabricated on a 10 μm ~$1\times10^{16}$ cm$^{-3}$ (001) HVPE epilayer from NCT Japan and an 8.8 μm ~$1\times10^{17}$ cm$^{-3}$ epitaxial film grown at UCSB via MOCVD on co-loaded (010), (110), and (011) Sn-doped substrates. MOCVD samples were grown with the same parameters as previous with higher doping.[44] Fabrication of the circular SBDs began with a Ni/Au/Ni stack deposited via e-beam evaporation and patterned using optical lithography and lift-off. Next, a 120 nm TiO$_2$/Al$_2$O$_3$ nanolaminate field plate dielectric was deposited using a plasma atomic layer deposition (ALD) process at 300 °C similar to our previous work.[44,54] After ALD, a 200/30 W ICP/RF bias etch was performed using BCl$_3$ gas to etch through the ALD dielectric and open the Ni/Au/Ni anodes. On areas with no anode metal, the ICP etch proceeded ~40 nm into the β-Ga$_2$O$_3$ layer below, damaging the material. The final step was another Ni/Au metal deposition to make circular anode contacts on the ICP damaged areas as well as create the field plate overlap on the metal-first SBDs. The final device structures for both the reference field-plated SBDs (FP-SBDs) and the ICP damaged SBDs are shown in FIG. 3(a).

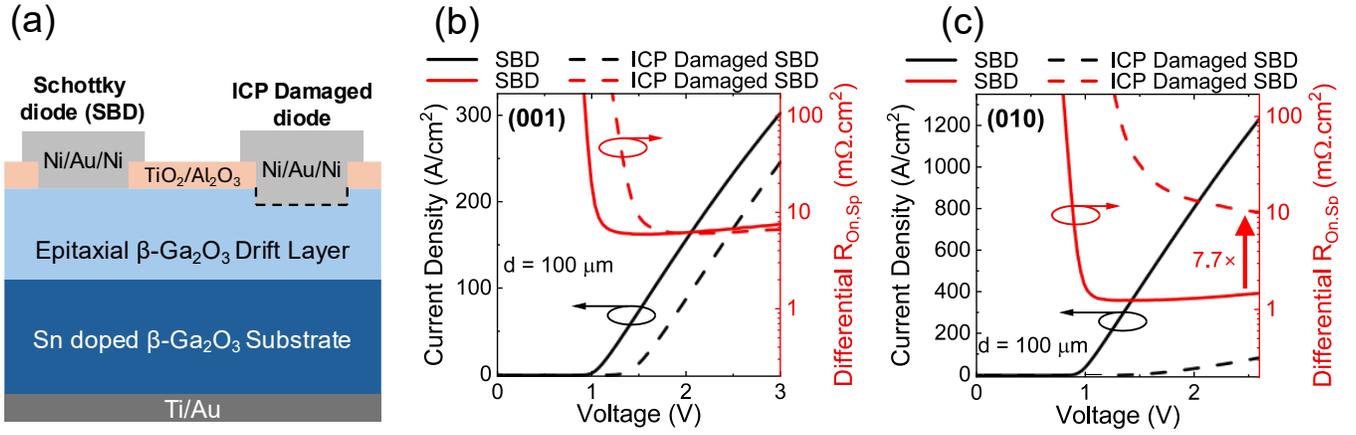

FIG. 3. (a) Device schematic showing reference metal-first FP-SBDs and ICP damaged SBDs co-fabricated on thick (001) and (010) $\beta$-$Ga_2O_3$ drift layers. (b) J-V and differential $R_{on,sp}$ plots for (001) SBD and ICP damaged SBDs (c) J-V and differential $R_{on,sp}$ plots for (010) SBD and ICP damaged SBDs, showing a 7.7× increase in $R_{on,sp}$ for ICP damaged (010) devices.

J-V measurements in FIG. 3(b) show that for the (001) material, the reference and damaged SBDs exhibit a shift in $V_{on}$ from 1.03 V to 1.45 V respectively, likely indicating an increase in the Schottky barrier height after ICP etching. Despite the change in $V_{on}$, the $R_{on,sp}$ value varies minimally, going from 6.01 to 6.11 m$\Omega$.cm$^2$ for the SBD and ICP damaged SBD, respectively. The change in the $V_{on}$ but constant nature of the $R_{on,sp}$ value after ICP etching suggests that the etching damage is constrained near the surface for the (001) material and does not affect the bulk of the epitaxial layer. Previous studies have shown an increase in the $R_{on,sp}$ from ICP etch damage in (001) SBDs.[15] However, our study uses a lower ICP and RF bias power (200/30 W), indicating that reducing the ion energy could be an effective way to minimize damage for (001) SBDs. The J-V for the (010) material is depicted in FIG. 3(c) and shows a significant decrease in the current density after ICP etching and a positive shift in the $V_{on}$ from 0.95 V to ~1.35 V for the reference SBD and damaged (010) SBDs, respectively. The $R_{on,sp}$ is significantly higher for the ICP damaged SBDs, with the value going from 1.27 to 9.73 m$\Omega$.cm$^2$ after the sample was exposed to the 30/200 W $BCl_3$ plasma. This represents a 7.7× increase in $R_{on,sp}$ and is comparable to the 9.4× increase in $R_{on,sp}$ observed from sputtered $NiO_x$ damage in FIG. 1(c). This further confirms that the energetic ion processes are the culprit for increased device $R_{on,sp}$ and that the $\beta$-$Ga_2O_3$ crystal experiences anisotropic damage, with the (010) orientation being highly susceptible to damage where the (001) plane is more robust.

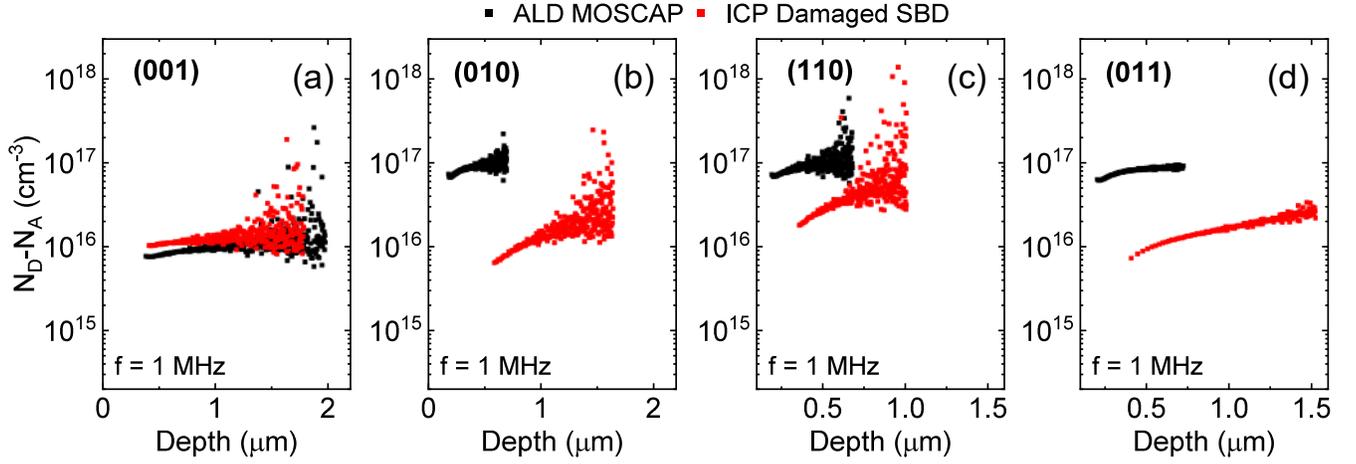

FIG. 4. $N_D - N_A$ vs. depth profiles of reference MOSCAPs and ICP damaged SBDs on a 10 μm thick $1\times10^{16}$ cm$^{-3}$ doped (a) (001) film and on 8.8 μm thick $1\times10^{17}$ cm$^{-3}$ doped (b) (010), (c) (110), and (d) (011) epitaxial layers. Depletion of $N_D - N_A$ from ICP damage is observed in all epitaxial layers except for the (001) orientation.

For C-V measurements, MOSCAPs were created by lithographically patterning Ni/Au circular anodes onto the 120 nm TiO$_2$/Al$_2$O$_3$ ALD nanolaminate dielectric. The ALD MOSCAPs produced identical $N_D - N_A$ concentrations as the metal first reference SBDs and minimized C-V leakage current. The full comparison of C-V $N_D - N_A$ profiles for the four crystal orientations is depicted in FIG. 4. ICP damaged diodes on the (001) oriented epilayer (FIG. 4(a)) exhibited a slight increase in charge with respect to the reference ALD MOSCAPs. The increase in charge is potentially due to lateral charge variation across the sample. Minimal reduction in $N_D - N_A$ for the (001) epilayer is consistent with the negligible change in R$_{on,sp}$ observed from J-V measurements (FIG. 3(b)). In contrast, the (010), (110), and (011) orientations (FIG. 4(b),(c),(d)) all show a reduction in $N_D - N_A$ (~85% reduction for each at zero bias) for the ICP damaged SBDs when compared to the ALD MOSCAPs. The reduction in $N_D - N_A$ for the ICP damaged diodes on (010) epilayers agrees with the increase in R$_{on,sp}$ from J-V results from FIG. 3(c). Additionally, the reduction of zero-bias $N_D - N_A$ for the ICP damaged (010) epilayer (91%) is comparable to the reduction seen from sputter damage (89%), further solidifying that (010) is very susceptible to ion damage. A complete comparison of the changes in R$_{on,sp}$ and zero bias $N_D - N_A$ are shown in Table I.

Table I. Summary of damage for various crystal orientations and fabrication processes

| β-Ga$_2$O$_3$ Crystal Orientation | Damaging Process | % change in R$_{on,sp}$ | % change in $N_D - N_A$ at zero Bias |
|---|---|---|---|
| (001) | Sputter NiO$_x$ | +6.91% | -0.86% |
| (001) | ICP | +1.64% | +34.44% |
| (010) | Sputter NiO$_x$ | +842.33% | -85.42% |
| (010) | Sputter SiO$_2$ | N/A | -88.78% |
| (010) | ICP | +666.48% | -91.00% |
| (110) | ICP | +539.18% | -75.07% |
| (011) | ICP | +21.26% | -88.30% |

A possible explanation for why (010), (110), and (011) are more susceptible to ion damage is seen in FIG. 5. The first row FIG. 5(a-d) shows VESTA[55] plots of the various orientations, which are one unit cell thick into the page. Additionally, the second row FIG. 5(e-h) depicts the same crystal orientations, but 4 unit cells thick into the page. Clearly, only the (010) orientation has well defined open channels as expected because the unique b-axis [010] orientation is co-axial with the two-fold axis and perpendicular to the one mirror plane in the crystal ($\frac{2}{m}$ point group). Conversely, the normals to the (001), (110) and (011) planes (reciprocal lattice vectors $\boldsymbol{g}_{001}, \boldsymbol{g}_{110},$ and $\boldsymbol{g}_{011}$) do not coincide with any symmetry elements. Thus, somewhat expectedly, the (001), (110) and (011) crystal orientations do not show clear ion channels (FIG. 5(e, g, h)). Open material channels, like those present in (010) β-Ga$_2$O$_3$, have been shown to increase the penetration depth of incident ions before encountering a scattering event.[56,57] This channeling effect could help explain why ion damage is observed deeper into the (010) film than the (001). Additionally, as the (110) and (011) planes are only 14.34° and 28.35° rotated from (010) respectively, as seen in FIG. 5(i, j), the open ion channeling pathways along [010] can be accessed easily from scattered ions, explaining why the (110) and (011) epitaxial layers also see significant ion damage.

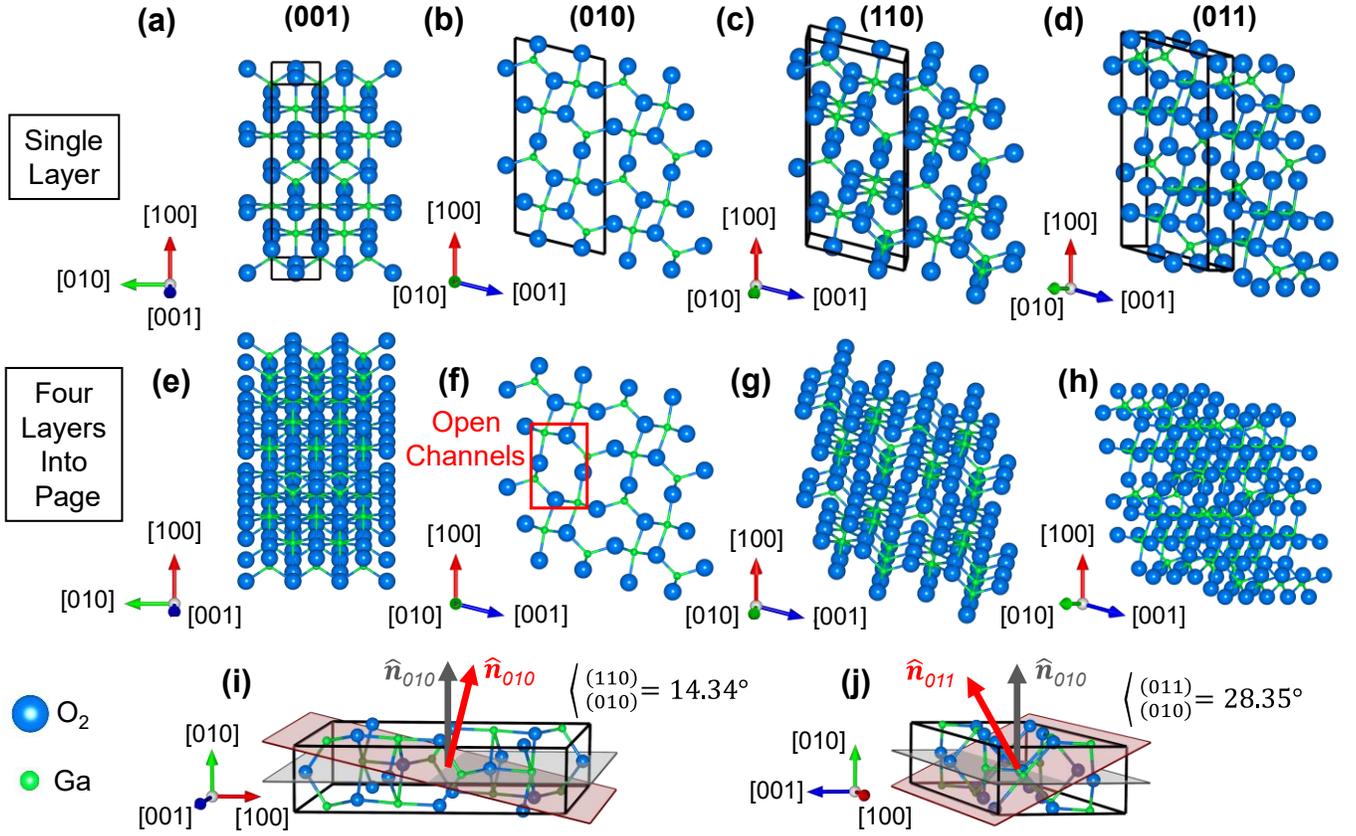

FIG. 5. β-Ga$_2$O$_3$ crystal structure models for the (a, e) (001), (b, f) (010), (c, g) (110), and (d, h) (011) planes, showing the large channeling pathways present in (010) plane when compared to the (001) plane. Row (a-d) depicts a crystal structure 1 unit cell thick, whereas row (e-h) depicts a crystal structure 4 unit cells thick into the page. The (110) and (011) planes are rotated from the (010) plane by an angle of (i) 14.34° and (j) 28.35° respectively, with the normal vectors $\hat{n} = \frac{g_{hkl}}{|g_{hkl}|}$ shown.

An important observation from FIG. 2(c) was that reduction in $N_D - N_A$ was observed as deep as 11.5 μm into the material. Since the observed damage is deeper than expected from ion damage alone,[56] a possible explanation is the combination of both ion channeling and point defect diffusion along the [010] direction. It is also likely that the Fermi level in the ion damaged top region moves towards the valence band, leading to enhanced charge depletion in the low-doped drift layers. Regardless, it is clear from C-V measurements that there is compensation of donors in the [010] direction from an increase in acceptor density $N_A$, but whether that comes from acceptor species,[58–63] gallium vacancies (V$_{Ga}$),[64–66] or another acceptor source remains unclear. Thus, it is hypothesized that the observed μm-scale damage is caused by energetic ions scattering into the open channels in the [010] direction and creating charge-compensating acceptors in the epitaxial layer. The compensating point defects could then diffuse further into the β-Ga$_2$O$_3$ layer along the [010] direction during the energetic etching/sputtering processes. However, the detailed mechanism for such long-range charge depletion needs further investigation.

In conclusion, we report on micron-scale charge depletion in (010), (110), and (011) $\beta$-Ga$_2$O$_3$ epitaxial layers $\beta$-Ga$_2$O$_3$ epitaxial layers of up to 11.5 µm due to ion damage from sputtering and ICP etching processes. Additionally, we report that the (001) epitaxial layers were more robust to the sputtering and ICP ion damage and showed minimal charge depletion. The anisotropic removal of charge was observed by comparing sputtered NiO$_x$ p-n HJDs and ICP damaged SBDs with low-damage reference SBDs. The resulting comparison for (010) epilayers showed a 9.4× and 7.7× increase in the R$_{on,sp}$ for the sputter-damaged NiO$_x$ HJDs and ICP damaged SBDs respectively. Additionally, an 85%, 89%, and 91% reduction in $N_D - N_A$ at zero bias was observed for the sputtered NiO$_x$ HJDs, sputtered SiO$_2$ MOSCAPs, and ICP damaged SBDs on the (010) films respectively. In contrast, the HJDs and ICP damaged diodes on (001) showed little change in R$_{on,sp}$ or $N_D - N_A$. Additionally, diodes fabricated on ICP damaged (110) and (011) epilayers saw a 75% and 88% reduction in $N_D - N_A$ respectively, indicating they are also susceptible to ion damage. The damage observed in the (010), (110), and (011) diodes is potentially due to ions entering into the crystal along the large open channels in the [010] direction and creating point defects. These defects could then diffuse deeper into the material during the energetic sputter and ICP processes, depleting charge. Thus, it is imperative that high-energy ion processes such as sputtering or ICP etching be avoided, minimized, or replaced with low-damage processes on the exposed (010) plane when processing $\beta$-Ga$_2$O$_3$ devices.

## ACKNOWLEDGMENTS

The authors acknowledge funding from the ARPA-E ULTRAFAST program (DE-AR0001824) and Coherent / II-VI Foundation Block Gift Program. A portion of this work was performed at the UCSB Nanofabrication Facility, an open access laboratory. Carl Peterson would like to thank Steve Rebollo for his helpful discussions on crystal orientations and ion damage.

## DATA AVAILABILITY

The data that support the findings of this study are available from the corresponding author upon reasonable request.

# REFERENCES

[1] A.J. Green, J. Speck, G. Xing, P. Moens, F. Allerstam, K. Gumaelius, T. Neyer, A. Arias-Purdue, V. Mehrotra, A. Kuramata, K. Sasaki, S. Watanabe, K. Koshi, J. Blevins, O. Bierwagen, S. Krishnamoorthy, K. Leedy, A.R. Arehart, A.T. Neal, S. Mou, S.A. Ringel, A. Kumar, A. Sharma, K. Ghosh, U. Singisetti, W. Li, K. Chabak, K. Liddy, A. Islam, S. Rajan, S. Graham, S. Choi, Z. Cheng, and M. Higashiwaki, "β-Gallium oxide power electronics," APL Materials **10**(2), 029201 (2022).

[2] J.D. Blevins, K. Stevens, A. Lindsey, G. Foundos, and L. Sande, "Development of Large Diameter Semi-Insulating Gallium Oxide ($Ga_2O_3$) Substrates," IEEE Transactions on Semiconductor Manufacturing **32**(4), 466–472 (2019).

[3] Z. Galazka, K. Irmscher, R. Uecker, R. Bertram, M. Pietsch, A. Kwasniewski, M. Naumann, T. Schulz, R. Schewski, D. Klimm, and M. Bickermann, "On the bulk β-$Ga_2O_3$ single crystals grown by the Czochralski method," Journal of Crystal Growth **404**, 184–191 (2014).

[4] Z. Galazka, "Growth of bulk *β*-$Ga_2O_3$ single crystals by the Czochralski method," Journal of Applied Physics **131**(3), 031103 (2022).

[5] K. Hoshikawa, E. Ohba, T. Kobayashi, J. Yanagisawa, C. Miyagawa, and Y. Nakamura, "Growth of β-$Ga_2O_3$ single crystals using vertical Bridgman method in ambient air," Journal of Crystal Growth **447**, 36–41 (2016).

[6] H. Aida, K. Nishiguchi, H. Takeda, N. Aota, K. Sunakawa, and Y. Yaguchi, "Growth of β-$Ga_2O_3$ Single Crystals by the Edge-Defined, Film Fed Growth Method," Jpn. J. Appl. Phys. **47**(11R), 8506 (2008).

[7] A. Kuramata, K. Koshi, S. Watanabe, Y. Yamaoka, T. Masui, and S. Yamakoshi, "High-quality β-$Ga_2O_3$ single crystals grown by edge-defined film-fed growth," Jpn. J. Appl. Phys. **55**(12), 1202A2 (2016).

[8] A. Yoshikawa, V. Kochurikhin, T. Tomida, I. Takahashi, K. Kamada, Y. Shoji, and K. Kakimoto, "Growth of bulk *β*-$Ga_2O_3$ crystals from melt without precious-metal crucible by pulling from a cold container," Sci Rep **14**(1), 14881 (2024).

[9] C. Peterson, A. Bhattacharyya, K. Chanchaiworawit, R. Kahler, S. Roy, Y. Liu, S. Rebollo, A. Kallistova, T.E. Mates, and S. Krishnamoorthy, "200 $cm^2$/Vs electron mobility and controlled low $10^{15}$ $cm^{-3}$ Si doping in (010) β-$Ga_2O_3$ epitaxial drift layers," Applied Physics Letters **125**(18), 182103 (2024).

[10] Y. Zhang, and J.S. Speck, "Importance of shallow hydrogenic dopants and material purity of ultra-wide bandgap semiconductors for vertical power electron devices," Semicond. Sci. Technol. **35**(12), 125018 (2020).

[11] M. Higashiwaki, K. Sasaki, A. Kuramata, T. Masui, and S. Yamakoshi, "Gallium oxide ($Ga_2O_3$) metal-semiconductor field-effect transistors on single-crystal β-$Ga_2O_3$ (010) substrates," Appl. Phys. Lett. **100**(1), 013504 (2012).

[12] Y. Liu, H. Wang, C. Peterson, J.S. Speck, C. Van de Walle, and S. Krishnamoorthy, "$Cr_2O_3$/*β*-$Ga_2O_3$ Heterojunction Diodes with Orientation-Dependent Breakdown Electric Field up to 12.9 MV/cm," arXiv:2511.20885, (2025).

[13] J.-S. Li, H.-H. Wan, C.-C. Chiang, T.J. Yoo, M.-H. Yu, F. Ren, H. Kim, Y.-T. Liao, and S.J. Pearton, "Breakdown up to 13.5 kV in NiO/*β*-$Ga_2O_3$ Vertical Heterojunction Rectifiers," ECS J. Solid State Sci. Technol. **13**(3), 035003 (2024).

[14] D. Wakimoto, C.-H. Lin, K. Ema, Y. Ueda, H. Miyamoto, K. Sasaki, and A. Kuramata, "A multi-fin normally-off *β*-$Ga_2O_3$ vertical transistor with a breakdown voltage exceeding 10 kV," Appl. Phys. Express **18**(10), 106502 (2025).

[15] C.-C. Chiang, X. Xia, J.-S. Li, F. Ren, and S.J. Pearton, "Ion energy dependence of dry etch damage depth in $Ga_2O_3$ Schottky rectifiers," Applied Surface Science **631**, 157489 (2023).

[16] S. Rebollo, T. Itoh, S. Krishnamoorthy, and J.S. Speck, "Heated-$H_3PO_4$ etching of (001) $\beta$-$Ga_2O_3$," Applied Physics Letters **125**(1), 012102 (2024).

[17] Y. Zhang, A. Mauze, and J.S. Speck, "Anisotropic etching of $\beta$-$Ga_2O_3$ using hot phosphoric acid," Appl. Phys. Lett. **115**(1), 013501 (2019).

[18] T. Oshima, and Y. Oshima, "Plasma-free dry etching of (001) $\beta$-$Ga_2O_3$ substrates by HCl gas," Applied Physics Letters **122**(16), 162102 (2023).

[19] Y. Oshima, E. Ahmadi, S. Kaun, F. Wu, and J.S. Speck, "Growth and etching characteristics of (001) $\beta$-$Ga_2O_3$ by plasma-assisted molecular beam epitaxy," Semicond. Sci. Technol. **33**(1), 015013 (2017).

[20] A. Katta, F. Alema, W. Brand, A. Gilankar, A. Osinsky, and N.K. Kalarickal, "Demonstration of MOCVD based in situ etching of β-$Ga_2O_3$ using TEGa," Journal of Applied Physics **135**(7), 075705 (2024).

[21] L. Meng, V.G. Thirupakuzi Vangipuram, D.S. Yu, C. Hu, and H. Zhao, "Metalorganic chemical vapor deposition in situ etching of $\beta$-$Ga_2O_3$ and $\beta$-$(Al_xGa_{1-x})_2O_3$," J. Vac. Sci. Technol. B **43**(5), 052201 (2025).

[22] S.A. Khan, A. Ibreljic, and A.F.M.A.U. Bhuiyan, "Plasma damage-free in situ etching of $\beta$-$Ga_2O_3$ using solid-source gallium in the LPCVD system," Appl. Phys. Lett. **127**(10), 102105 (2025).

[23] D. Kumar, and A. Verma, "Gallium flux induced etching of $\beta$-$Ga_2O_3$ in an LPCVD system," J. Vac. Sci. Technol. A **43**(5), 053405 (2025).

[24] Z. Hu, K. Nomoto, W. Li, N. Tanen, K. Sasaki, A. Kuramata, T. Nakamura, D. Jena, and H.G. Xing, "Enhancement-Mode $Ga_2O_3$ Vertical Transistors With Breakdown Voltage >1 kV," IEEE Electron Device Lett. **39**(6), 869–872 (2018).

[25] W. Li, K. Nomoto, Z. Hu, T. Nakamura, D. Jena, and H.G. Xing, "Single and multi-fin normally-off $Ga_2O_3$ vertical transistors with a breakdown voltage over 2.6 kV," in *2019 IEEE International Electron Devices Meeting (IEDM)*, (2019), p. 12.4.1-12.4.4.

[26] K. Tetzner, M. Klupsch, A. Popp, S. Bin Anooz, T.-S. Chou, Z. Galazka, K. Ickert, M. Matalla, R.-S. Unger, E.B. Treidel, M. Wolf, A. Trampert, J. Würfl, and O. Hilt, "Enhancement-mode vertical (100) $\beta$-$Ga_2O_3$ FinFETs with an average breakdown strength of 2.7 MV $cm^{-1}$," Jpn. J. Appl. Phys. **62**(SF), SF1010 (2023).

[27] S. Roy, C.N. Saha, C. Peterson, W.J. Mitchell, J.S. Speck, and S. Krishnamoorthy, "Multi-Fin $\beta$-$Ga_2O_3$ Vertical FinFET With Interfin Field Oxide Exhibiting a Breakdown Voltage of 1.8 kV and Power Figure of Merit of 1 $GW/cm^2$," IEEE Electron Device Letters **47**(2), 341–344 (2026).

[28] K.D. Chabak, N. Moser, A.J. Green, D.E. Walker Jr., S.E. Tetlak, E. Heller, A. Crespo, R. Fitch, J.P. McCandless, K. Leedy, M. Baldini, G. Wagner, Z. Galazka, X. Li, and G. Jessen, "Enhancement-mode $Ga_2O_3$ wrap-gate fin field-effect transistors on native (100) $\beta$-$Ga_2O_3$ substrate with high breakdown voltage," Applied Physics Letters **109**(21), 213501 (2016).

[29] Y. Zhang, A. Mauze, F. Alema, A. Osinsky, T. Itoh, and J.S. Speck, "$\beta$-$Ga_2O_3$ lateral transistors with high aspect ratio fin-shape channels," Jpn. J. Appl. Phys. **60**(1), 014001 (2020).

[30] A. Bhattacharyya, S. Roy, P. Ranga, C. Peterson, and S. Krishnamoorthy, "High-Mobility Tri-Gate β-$Ga_2O_3$ MESFETs With a Power Figure of Merit Over 0.9 $GW/cm^2$," IEEE Electron Device Letters **43**(10), 1637–1640 (2022).

[31] H. Liu, J. Li, Y. Lv, Y. Wang, X. Lu, S. Dun, T. Han, H. Guo, A. Bu, X. Ma, Z. Feng, and Y. Hao, "Improved electrical performance of lateral $\beta$-$Ga_2O_3$ MOSFETs utilizing slanted fin channel structure," Appl. Phys. Lett. **121**(20), 202101 (2022).

[32] N. Das, C. Gorsak, A. Kashyap, A. Gilankar, P. Tripathi, S.A. Paranjape, T. Thornton, H. Nair, and N. Kurian Kalarickal, "$\beta$-$Ga_2O_3$ Sub-Micron FinFETs With Si Delta-Doped Channel Modulating >3 × $10^{13}$ $cm^{-2}$ Charge Density," IEEE Electron Device Letters **47**(3), 466–469 (2026).

[33] K. Sasaki, D. Wakimoto, Q.T. Thieu, Y. Koishikawa, A. Kuramata, M. Higashiwaki, and S. Yamakoshi, "First Demonstration of Ga$_2$O$_3$ Trench MOS-Type Schottky Barrier Diodes," IEEE Electron Device Letters **38**(6), 783–785 (2017).

[34] W. Li, Z. Hu, K. Nomoto, R. Jinno, Z. Zhang, T.Q. Tu, K. Sasaki, A. Kuramata, D. Jena, and H.G. Xing, "2.44 kV Ga$_2$O$_3$ vertical trench Schottky barrier diodes with very low reverse leakage current," in *2018 IEEE International Electron Devices Meeting (IEDM)*, (2018), p. 8.5.1-8.5.4.

[35] S. Roy, A. Bhattacharyya, C. Peterson, and S. Krishnamoorthy, "β-Ga$_2$O$_3$ Lateral High-Permittivity Dielectric Superjunction Schottky Barrier Diode With 1.34 GW/cm$^2$ Power Figure of Merit," IEEE Electron Device Letters **43**(12), 2037–2040 (2022).

[36] S. Roy, B. Kostroun, J. Cooke, Y. Liu, A. Bhattacharyya, C. Peterson, B. Sensale-Rodriguez, and S. Krishnamoorthy, "Ultra-low reverse leakage in large area kilo-volt class β-Ga$_2$O$_3$ trench Schottky barrier diode with high-k dielectric RESURF," Applied Physics Letters **123**(24), 243502 (2023).

[37] S. Roy, B. Kostroun, Y. Liu, J. Cooke, A. Bhattacharyya, C. Peterson, B. Sensale-Rodriguez, and S. Krishnamoorthy, "Low Q$_C$V$_F$ 20 A/1.4 kV β-Ga$_2$O$_3$ Vertical Trench High-k RESURF Schottky Barrier Diode With Turn-On Voltage of 0.5 V," IEEE Electron Device Letters **45**(12), 2487–2490 (2024).

[38] C.N. Saha, S. Roy, Y. Liu, C. Peterson, and S. Krishnamoorthy, "2.34 kV β-Ga$_2$O$_3$ vertical trench RESURF Schottky barrier diode with sub-micron fin width," APL Electronic Devices **1**(4), 046125 (2025).

[39] G. Guo, X. Zhang, C. Zeng, D. Wei, D. Zhao, T. Chen, Z. Li, A. Luo, G. Yu, Y. Hu, Z. Zeng, B. Zhang, and X. Dai, "1.1 kV/0.72 GW/cm$^2$ β-Ga$_2$O$_3$ Fin-channel diode with ohmic contacts anode," J. Semicond. **47**(3), 032502 (2026).

[40] W. Li, K. Nomoto, Z. Hu, D. Jena, and H.G. Xing, "Fin-channel orientation dependence of forward conduction in kV-class Ga2O3 trench Schottky barrier diodes," Appl. Phys. Express **12**(6), 061007 (2019).

[41] S. Dhara, N.K. Kalarickal, A. Dheenan, C. Joishi, and S. Rajan, "β-Ga$_2$O$_3$ Schottky barrier diodes with 4.1 MV/cm field strength by deep plasma etching field-termination," Appl. Phys. Lett. **121**(20), 203501 (2022).

[42] H.-S. Kim, A.K. Bhat, M.D. Smith, and M. Kuball, "Turn-On Voltage Instability of β-Ga$_2$O$_3$ Trench Schottky Barrier Diodes With Different Fin Channel Orientations," IEEE Transactions on Electron Devices **71**(6), 3609–3613 (2024).

[43] S. Rebollo, W. Buchmaier, S. Krishnamoorthy, and J.S. Speck, "Dry etch damage anisotropy and damage mitigation using hot H$_3$PO$_4$ in (001) β-Ga$_2$O$_3$ Schottky diodes," APL Electronic Devices **2**(1), 016117 (2026).

[44] C. Peterson, C.N. Saha, R. Kahler, Y. Liu, A. Mattapalli, S. Roy, and S. Krishnamoorthy, "Kilovolt-class β-Ga$_2$O$_3$ field-plated Schottky barrier diodes with MOCVD-grown intentionally 10$^{15}$ cm$^{-3}$ doped drift layers," J. Appl. Phys. **138**(18), 185105 (2025).

[45] Y. Kokubun, S. Kubo, and S. Nakagomi, "All-oxide p–n heterojunction diodes comprising p-type NiO and n-type β-Ga$_2$O$_3$," Appl. Phys. Express **9**(9), 091101 (2016).

[46] X. Lu, X. Zhou, H. Jiang, K.W. Ng, Z. Chen, Y. Pei, K.M. Lau, and G. Wang, "1-kV Sputtered p-NiO/n-Ga$_2$O$_3$ Heterojunction Diodes With an Ultra-Low Leakage Current Below 1 uA/cm$^2$," IEEE Electron Device Letters **41**(3), 449–452 (2020).

[47] J. Zhang, P. Dong, K. Dang, Y. Zhang, Q. Yan, H. Xiang, J. Su, Z. Liu, M. Si, J. Gao, M. Kong, H. Zhou, and Y. Hao, "Ultra-wide bandgap semiconductor Ga$_2$O$_3$ power diodes," Nat Commun **13**(1), 3900 (2022).

[48] Y. Wang, H. Gong, Y. Lv, X. Fu, S. Dun, T. Han, H. Liu, X. Zhou, S. Liang, J. Ye, R. Zhang, A. Bu, S. Cai, and Z. Feng, "2.41 kV Vertical p-NiO/n-Ga$_2$O$_3$ Heterojunction Diodes With a Record

⁴⁸ Baliga's Figure-of-Merit of 5.18 GW/cm$^2$," IEEE Transactions on Power Electronics **37**(4), 3743–3746 (2022).

⁴⁹ H. Gong, F. Zhou, M. Xiao, Z. Yang, F.-F. Ren, S. Gu, H. Lu, R. Zhang, Y. Zhang, and J. Ye, "Enhanced avalanche (2.1 kV, 83 A) in NiO/Ga$_2$O$_3$ heterojunction by edge termination optimization," IEEE Electron Device Letters **45**(8), 1421–1424 (2024).

⁵⁰ Y. Liu, S.M.W. Witsell, J.F. Conley Jr., and S. Krishnamoorthy, "Orientation-dependent β-Ga$_2$O$_3$ heterojunction diode with atomic layer deposition (ALD) NiO," Appl. Phys. Lett. **127**(12), 122109 (2025).

⁵¹ Y. Liu, S. Roy, C. Peterson, A. Bhattacharyya, and S. Krishnamoorthy, "Ultra-low reverse leakage NiO$_x$/β-Ga$_2$O$_3$ heterojunction diode achieving breakdown voltage >3 kV with plasma etch field-termination," AIP Advances **15**(1), 015114 (2025).

⁵² H. Gong, X. Yang, B. Wang, Z. Zhang, Q. Yuchi, Z. Yang, M. Porter, H. Cui, Y. Qin, R. Zhang, H. Wang, D. Dong, J. Ye, G.-Q. Lu, and Y. Zhang, "A megawatt ultra-wide bandgap semiconductor module for pulsed power electronics," Nat Commun, (2026).

⁵³ M.A. Porter, Y. Ma, Y. Qin, and Y. Zhang, "P-Type Doping Control of Magnetron Sputtered NiO for High Voltage UWBG Device Structures," in *2023 IEEE 10th Workshop on Wide Bandgap Power Devices & Applications (WiPDA)*, (2023), pp. 1–7.

⁵⁴ C.N. Saha, S. Roy, Y. Liu, C. Peterson, and S. Krishnamoorthy, "2.34 kV β-Ga$_2$O$_3$ vertical trench RESURF Schottky barrier diode with sub-micron fin width," APL Electronic Devices **1**(4), 046125 (2025).

⁵⁵ K. Momma, and F. Izumi, "VESTA 3 for three-dimensional visualization of crystal, volumetric and morphology data," J Appl Cryst **44**(6), 1272–1276 (2011).

⁵⁶ T. Liu, Z. Li, J. Zhao, X. Fei, J. Feng, Y. Zuo, M. Hua, Y. Guo, S. Liu, and Z. Zhang, "Orientation-dependent surface radiation damage in β-Ga$_2$O$_3$ explored by atomistic simulations," Acta Materialia **300**, 121484 (2025).

⁵⁷ C.-H. Chen, D.L. Green, and E.L. Hu, "Diffusion and channeling of low-energy ions: The mechanism of ion damage," J. Vac. Sci. Technol. B **13**(6), 2355–2359 (1995).

⁵⁸ M.H. Wong, C.-H. Lin, A. Kuramata, S. Yamakoshi, H. Murakami, Y. Kumagai, and M. Higashiwaki, "Acceptor doping of β-Ga$_2$O$_3$ by Mg and N ion implantations," Appl. Phys. Lett. **113**(10), 102103 (2018).

⁵⁹ A.T. Neal, S. Mou, S. Rafique, H. Zhao, E. Ahmadi, J.S. Speck, K.T. Stevens, J.D. Blevins, D.B. Thomson, N. Moser, K.D. Chabak, and G.H. Jessen, "Donors and deep acceptors in β-Ga$_2$O$_3$," Appl. Phys. Lett. **113**(6), 062101 (2018).

⁶⁰ H. Peelaers, J.L. Lyons, J.B. Varley, and C.G. Van de Walle, "Deep acceptors and their diffusion in Ga$_2$O$_3$," APL Mater. **7**(2), 022519 (2019).

⁶¹ A. Mauze, Y. Zhang, T. Itoh, T.E. Mates, H. Peelaers, C.G. Van de Walle, and J.S. Speck, "Mg doping and diffusion in (010) β-Ga$_2$O$_3$ films grown by plasma-assisted molecular beam epitaxy," Journal of Applied Physics **130**(23), 235301 (2021).

⁶² C. Lee, M.A. Scarpulla, E. Ertekin, and J.B. Varley, "Diffusion of acceptor dopants in monoclinic β-Ga$_2$O$_3$," Phys. Rev. Mater. **9**(9), 094602 (2025).

⁶³ R. Kahler, C. Peterson, and S. Krishnamoorthy, "Realization of Insulating Buffer Layers via MOCVD-Grown Nitrogen-Doped (010) β-Ga$_2$O$_3$," arXiv:2512.20989, (2025).

⁶⁴ E. Korhonen, F. Tuomisto, D. Gogova, G. Wagner, M. Baldini, Z. Galazka, R. Schewski, and M. Albrecht, "Electrical compensation by Ga vacancies in Ga$_2$O$_3$ thin films," Appl. Phys. Lett. **106**(24), 242103 (2015).

⁶⁵ P. Deák, Q. Duy Ho, F. Seemann, B. Aradi, M. Lorke, and T. Frauenheim, "Choosing the correct hybrid for defect calculations: A case study on intrinsic carrier trapping in β-Ga$_2$O$_3$," Phys. Rev. B **95**(7), 075208 (2017).

[66] A. Kyrtsos, M. Matsubara, and E. Bellotti, "Migration mechanisms and diffusion barriers of vacancies in $Ga_2O_3$," Phys. Rev. B **95**(24), 245202 (2017).